\documentclass[11pt]{article}
\usepackage[dvips]{graphicx}
\begin{document}
\title{Fluctuation Dissipation Relation for a Langevin Model with Multiplicative Noise} 
\author{
Hidetsugu Sakaguchi\\
Department of Applied Science for Electronics and Materials,\\ 
Interdisciplinary Graduate School of Engineering Sciences,\\
 Kyushu University, Kasuga, Fukuoka 816-8580, Japan}
\maketitle
\begin{abstract}
A random multiplicative process with additive noise is described by a 
Langevin equation.   
We show that the
fluctuation-dissipation relation is 
satisfied in the Langevin model, if the noise strength  is not so strong.\\
\\
keywords\\
fluctuation-dissipation theorem, Langevin equation, multiplicative noise, Levy flight.
\end{abstract}

\section{Introduction and a Langevin equation}
The Einstein relation and the fluctuation-dissipation theorem are 
important relations in the nonequilibrium statistical physics.~\cite{rf:1,rf:2} There are many attemps to extend those relations in systems far from equilibrium~\cite{rf:3,rf:4}.
Langevin type equations have been recently studied to describe some nonequilibrium systems such as 
ratchet models~\cite{rf:5,rf:6,rf:7} 
and random multiplicative processes~\cite{rf:8,rf:9,rf:10}.
The random multiplicative processes are used to describe large fluctuations 
in chaotic dynamical systems and economic activity.
The Langevin equation with multiplicative noise generates a power-law distribution in contrast to the Gaussian distribution in thermal equilibrium. 
We study  the fluctuation-dissipation relation in this Langevin equation.   

We study a Langevin equation for a random multiplicative process with additive noise:
\begin{equation}
\frac{dy}{dt}=(-\gamma+\xi(t))y(t)+\eta(t),
\end{equation}
where $y(t)$ is a stochastic variable, $\gamma$ is a damping constant, $\xi(t)$ denotes multiplicative noise and $\eta(t)$ addtive noise. The noises are assumed to be Gaussian-white noises satisfying
\[\langle \xi(t)\xi(t^{\prime})\rangle=2D\delta(t-t^{\prime}),\;\; \langle \eta(t)\eta(t^{\prime})\rangle=2T\delta(t-t^{\prime}).\]
If $D$ is zero, this process is equivalent to the Ornstein-Ulenbeck process with temperature $T_a$ where $k_BT_a=T/\gamma$. 
The fluctuation-dissipation theorem and the Einstein relation are explicitly solved in this Ornstein-Ulenbeck process.  We will show a generalized fluctuation-dissipation relation for Eq.~(1) with nonzero $D$. 

\section{Time correlation and the fluctuation-dissipation relation}
The Fokker-Planck equation for the Langevin equation (1) is written as
\begin{equation}
\frac{\partial P}{\partial t}=-\frac{\partial}{\partial y}(-\gamma y P)+D\frac{\partial}{\partial y}y\left(\frac{\partial }{\partial y}yP\right )+T\frac{\partial^2 P}{\partial y^2},
\end{equation}
where $P(y,t)$ is the probability density function for $y$.
The stationary solution $P_0(y)$ of Eq.~(2) satisfies
\begin{equation}
\frac{\partial P_0}{\partial y}=-\frac{(\gamma+D)y}{T+Dy^2}P_0.
\end{equation}
The stationary distribution has a form
\begin{equation}
P_0(y)=\frac{c_0}{(T+Dy^2)^{(\gamma+D)/(2D)}},
\end{equation}
where $c_0$ is a normalization constant. The normalization constant is given by 
\[c_0=\frac{T^{\epsilon/2}D^{1/2}\Gamma(1/2+\epsilon/2)}{\Gamma(1/2)\Gamma(\epsilon/2)},\]
where $\epsilon=\gamma/D$, and $\Gamma(x)$ is the $\Gamma$-function. 
The variance of $y$ can be calculated as
\[\langle y^2\rangle=\int_{-\infty}^{\infty}y^2P_0(y)dy=\frac{T}{D}\frac{\Gamma(3/2)\Gamma(\epsilon/2-1)}{\Gamma(1/2)\Gamma(\epsilon/2)}=\frac{T}{\gamma-2D},\]
for $D<\gamma/2$. The variance diverges for $D>\gamma/2$, since $P_0(y)$ in Eq.~(4) decays more slowly than $|y^{-3}|$ for $y\rightarrow \pm\infty$.
We have performed a numerical simulation of the Langevin model (1) using the Heun method with timestep $\Delta t=10^{-3}$ or $10^{-4}$ to check the theoretical results. Figure 1 displays the numerically obtained probability distribution function (solid curve) and the distribution $P_0(y)$ in Eq.~(4) (dashed curve) for $\gamma=1,\;T=1/2,\;D=3/4$ in the logarithmic scale. The tail of the probability distribution obeys the power law with exponent $1+\epsilon=7/3$. 
\begin{figure}[htb]
\begin{center}
\includegraphics[width=7cm]{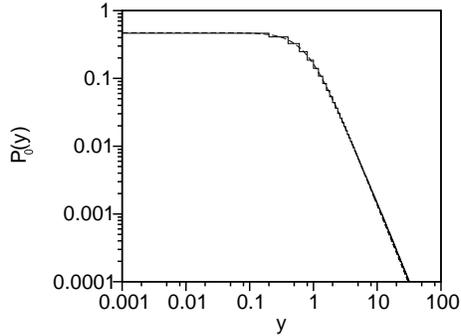}
\caption{Stationary probability distribution $P_0(y)$ for Eq.~(1)   
at $\gamma=1,T=1/2$ and $D=3/4$. The histogram is the numerically obtained distribution and the dashed line is the distribution function (4).} 
\label{fig:1} 
\end{center}
\end{figure} 

The Fokker-Planck equation and the stationary distribution are related to the Tsallis statistics~\cite{rf:11}. If $E(y)=1/2(\gamma+D)y^2$ is 
interpreted as generalized energy and $q$ is assumed to be $1+2D/(\gamma+D)$, 
the stationary distribution is rewritten as  $P_0(E)\propto 1/\{1+(q-1)E/T\}^{q-1}$, which has a form of the equilibrium distribution in the Tsallis statistics. If $D=0$, $q$ becomes 1 and the Boltzmann-Gibbs statistics is recovered.
The generalized free energy $F_q$ is defined in the Tsallis statistics as 
\[F_q=U_q-TS_q=\int_{-\infty}^{\infty} E(y)P^q(y)dy-T\frac{1-\int_{-\infty}^{\infty} P^q(y)dy}{q-1}dy,\]
where $U_q$ is the generalized internal energy and $S_q$ is the Tsallis entropy. 
The generalized free energy decreases monotonically in the time evolution of the Fokker-Planck equation, since 
\begin{eqnarray}
\frac{dF_q}{dt}&=&\int_{-\infty}^{\infty}\frac{1}{2}(\gamma+D)y^2qP^{q-1}\frac{\partial P}{\partial t}dy+\frac{T}{q-1}\int_{-\infty}^{\infty} qP^{q-1}\frac{\partial P}{\partial t}dy,\nonumber\\
&=&-q\int_{-\infty}^{\infty} P^{q-2}\{(T+Dy^2)\frac{\partial P}{\partial y}+(\gamma+D)yP\}^2dy\le 0.
\end{eqnarray}
This is a kind of the H-theorem in our stochastic process~\cite{rf:12,rf:13}.
The generalized free energy takes a minimum at the stationary distribution 
which satisfies Eq.~(3).

The solution of the linear equation (1) is explicitly given by
\begin{equation}
y(t)=y(0)e^{-\gamma t+\Xi(t)}+\int_0^te^{\gamma(t^{\prime}-t)-\Xi(t^{\prime})+\Xi(t)}\eta(t^{\prime})dt^{\prime},
\end{equation}
where 
$\Xi(t)=\int_0^t\xi(t^{\prime\prime})dt^{\prime\prime}$.
The time correlation $c(t)=\langle y(0)y(t)\rangle$ is given by
\begin{equation}
C(t)=\langle y(0)y(t)\rangle=\langle y(0)^2\rangle \langle e^{-\gamma t+\Xi(t)}\rangle,
\end{equation}
since there is no correlation between $y(0)$ and $\eta(t)$ and the average value of $\eta(t)$ is zero. The probability distribution of $\Xi(t)$ is the Gaussian distribution of variance $2Dt$. The time correlation is therefore given by 
\begin{equation}
C(t)=\frac{T}{\gamma-2D}e^{-(\gamma-D)t}.
\end{equation}
The time correlation decays exponentially and 
the decay constant is $1/(\gamma-D)$. The decay constant diverges at $D=\gamma$. For $\gamma/2<D<\gamma$, the variance of $y$ diverges but the decay constant does not diverge.
\begin{figure}[htb]
\begin{center}
\includegraphics[width=7cm]{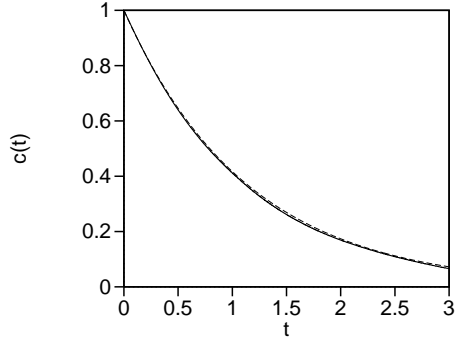}
\caption{Time correlation function $c(t)$ for Eq.~(1) at $\gamma=1,T=1/2$ and $D=1/8$.  The solid curve denotes a numerically obtained time correlation and the dashed curve denotes $e^{-(\gamma-D)t}$.} 
\label{fig:2} 
\end{center}
\end{figure} 
Figure 2 compares the numerically calculated correlation function $c(t)=\langle y(0)y(t)\rangle /\langle y(0)^2\rangle $ (solid curve) with the exponential decay $e^{-(\gamma-D)t}$ (dashed curve) 
at $\gamma=1,\;T=1/2,\;D=1/8$.

We have also investigated the response function for an external periodic force 
to Eq.~(1).  The model equation is given by 
\begin{equation}
\frac{dy}{dt}=(-\gamma+\xi(t))y(t)+f\exp(i\omega t)+\eta(t),
\end{equation}
where $f$ is the strength and $\omega$ is the frequency of the external force. The probability distribution function $P_f$ for the static force with $\omega=0$ 
obeys the Fokker-Planck equation including the external force term as
\begin{equation}
\frac{\partial P_f}{\partial t}=-\frac{\partial}{\partial y}\{(-\gamma y+f) P_f\}+D\frac{\partial}{\partial y}y\left(\frac{\partial }{\partial y}yP_f\right )+T\frac{\partial^2 P_f}{\partial y^2}.
\end{equation}
The stationary solution to this Fokker-Planck equation is given by
\begin{equation}
P_{f0}=c_{f0}\frac{\exp\{f\tan^{-1}(\sqrt{D/T}y)/\sqrt{TD}\}}{(T+Dy^2)^{(\gamma+D)/(2D)}},
\end{equation}
where $c_{f0}$ is a normalization constant.
The average value $\langle y\rangle$   for sufficiently small $f$ can be calculated as 
\begin{eqnarray}
 \langle y\rangle&\sim &fc_0\int_{-\infty}^{\infty}dy\frac{y\tan^{-1}(\sqrt{D/T}y)}{\sqrt{TD}(T+Dy^2)^{(\gamma+D)/(2D)}}\nonumber\\
&=&fc_0T^{-\epsilon/2}D^{-1/2}\frac{2}{\gamma-D}\int_0^{\pi/2}(\cos\theta)^{\epsilon-1}d\theta=\frac{f}{\gamma-D},
\end{eqnarray}
since $c_{f0}\sim c_0$ for sufficiently small $f$.
The static response function is simply written as $\chi_{st}=(\gamma-D)^{-1}$.

The general response function for the periodic force $f\exp(i\omega t)$  is expressed as 
\[\langle y(t)\rangle =(\chi^{\prime}(\omega)+i\chi^{\prime\prime}(\omega))fe^{i\omega t},\]
where $\chi(\omega)=\chi^{\prime}(\omega)+i\chi^{\prime\prime}(\omega)$ is called the complex admittance.
The Langevin equation is a linear equation and therefore the linear response 
relation is naturally expected. The fluctuation-dissipation relation 
can be generally expressed  as
\begin{equation}
G(\omega)=\frac{-2\chi^{\prime\prime}(\omega)}{\chi_{st}\omega}\langle y^2\rangle,\end{equation}
where $G(\omega)$ is the Fourier transform of the time correlation function $C(t)$. The standard fluctuation-dissipation theorem is obtained, if the thermal equilibrium condition $\langle y^2\rangle=k_BT_a\chi_{st}$ 
is further assumed.  
The time correlation function and the static response function $\chi_{st}$ have been already calculated and then the imaginary part of the response 
function is expected as 
\begin{equation}
\chi^{\prime\prime}(\omega)=\frac{-\omega}{(\gamma-D)^2+\omega^2}.
\end{equation}
\begin{figure}[htb]
\begin{center}
\includegraphics[width=7cm]{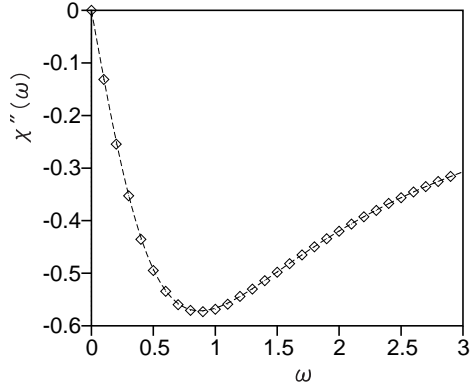}
\caption{Imaginary part $\chi^{\prime\prime}(\omega)$ of the complex admittance for $\gamma=1,T=1/2$ and $D=1/8$.
The numerical results are shown by points and the dashed curve denotes the function (14).} 
\label{fig:3} 
\end{center}
\end{figure}

On the other hand, 
the solution to the linear equation (9) is explicitly given by
\begin{equation}
y(t)=y(0)e^{-\gamma t+\Xi(t)}+\int_0^te^{\gamma(t^{\prime}-t)-\Xi(t^{\prime})+\Xi(t)}\{fe^{i\omega t^{\prime}}+\eta(t^{\prime})\}dt^{\prime}.
\end{equation}
The average time evolution of $y(t)$ after a long time is given by
\begin{equation}
\langle y(t)\rangle \sim \int_0^t\langle e^{\gamma(t^{\prime}-t)-\Xi(t^{\prime})+\Xi(t)}\rangle fe^{i\omega t^{\prime}}dt^{\prime}=\int_0^t e^{(\gamma-D)(t^{\prime}-t)}fe^{i\omega t^{\prime}}dt^{\prime}\sim \frac{fe^{i\omega t}}{\gamma-D+i\omega},
\end{equation}
where the relation $\langle \exp\{\Xi(t)-\Xi(t^{\prime})\}\rangle=\exp\{D(t-t^{\prime})\}$ is used, since the stochastic variable $\Xi(t)-\Xi(t^{\prime})$ obeys the Gaussian distribution with variance $2D(t-t^{\prime})$.
The complex admittance is calculated as 
\[\chi^{\prime}(\omega)=\frac{\gamma-D}{(\gamma-D)^2+\omega^2},\]
\begin{equation}
\chi^{\prime\prime}(\omega)=\frac{-\omega}{(\gamma-D)^2+\omega^2}.
\end{equation}
Equation (17) is equivalent to Eq.~(14), that is, it is shown that the fluctuation-dissipation relation is  satisfied  in our model equation.

We have numerically checked the fluctuation-dissipation relation.
We have calculated a Langevin equation with a sinusoidal force:
\begin{equation}
\frac{dy}{dt}=(-\gamma+\xi(t))y(t)+f\cos(i\omega t).
\end{equation}
We have neglected the additive noise to simplify the numerical simulation.
The imaginary part of the response function is numerically estimated from
\[\chi^{\prime\prime}(\omega)\sim\frac{-2}{fT_m}\int_0^{T_m}y(t)\sin(\omega t)dt,\]
where $T_m$ is a time interval to take a long-time average. The time interval $T_m=20000$ is used in our numerical simulation. 
Figure 3 displays the results for $\gamma=1\; D=1/8$ and $f=0.1$.
The points denote the numerically estimated values and the solid curve is the function (14).  We can see that the fluctuation-dissipation relation is satisfied in this random multiplicative process.

\section{Browinian motion and the Levy flight}
The Ornstein-Ulenbeck process is the simplest model for the Brownian motion. 
For the Brownian motion with velocity $y(t)$, the position $x(t)$ of the particle  obeys 
\begin{equation}
\frac{dx}{dt}=y(t). 
\end{equation}
It is a problem to find the diffusion coefficient of the Brownian particle.      The time correlation decays exponentially for $D<\gamma$. Then, the variance of the displacement $\Delta x(t)=x(t)-x(0)$ is calculated as
\begin{eqnarray}
\langle \Delta x(t)^2\rangle&=&\int_0^t\int_0^t\langle y(t^{\prime})y(t^{\prime\prime}\rangle dt^{\prime}dt^{\prime\prime}=2\langle y(0)^2\rangle\int_0^t\int_0^{t-t^{\prime}}c(t^{\prime\prime})dt^{\prime\prime}dt^{\prime}\nonumber\\
&\sim&\frac{2Tt}{(\gamma-D)(\gamma-2D)},
\end{eqnarray} 
for $D<\gamma/2$.
\begin{figure}[htb]
\begin{center}
\includegraphics[width=7cm]{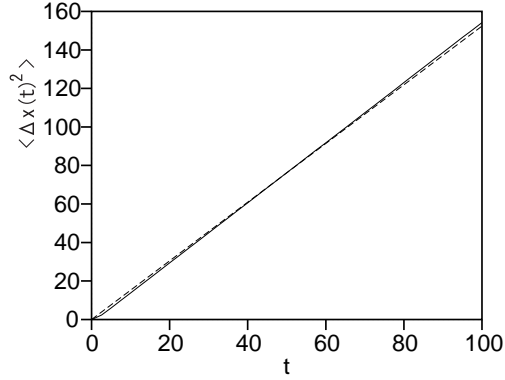}
\caption{Mean square $\langle \Delta x(t)^2\rangle$ of the displacement 
as a function of $t$ for Eq.~(1) and Eq.~(19) at $\gamma=1,T=1/2$ and $D=1/8$.
The numerical results are shown by the solid curve and the theoretical estimate (20) is shown by the dashed line.} 
\label{fig:4} 
\end{center}
\end{figure} 
\begin{figure}[htb]
\begin{center}
\includegraphics[width=7cm]{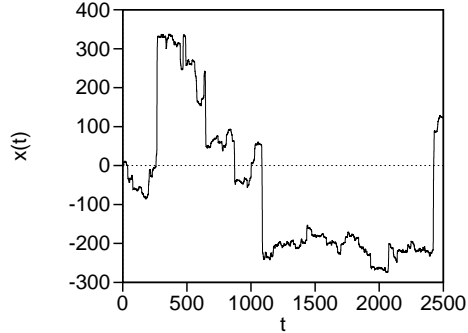}
\caption{Time evolution of $x(t)$ for Eq.~(1) and Eq.~(19) at $\gamma=1,T=1/2$ and $D=3/4$. The large fluctuation is characteristic of the Levy flight in one dimension.} 
\label{fig:5} 
\end{center}
\end{figure} 
The diffusion constant of the Brownian particle is written as $T/(\gamma-D)(\gamma-2D)$. This is a generalization of the Einstein relation between the diffusion constant and the mobility of the Brownian particle. Figure 4 displays $ \langle \Delta x(t)^2\rangle$ as a function of $t$ (solid curve) and the theoretical line (20) for
$\gamma=1,\;T=1/2$ and $D=1/8$. 
\begin{figure}[htb]
\begin{center}
\includegraphics[width=7cm]{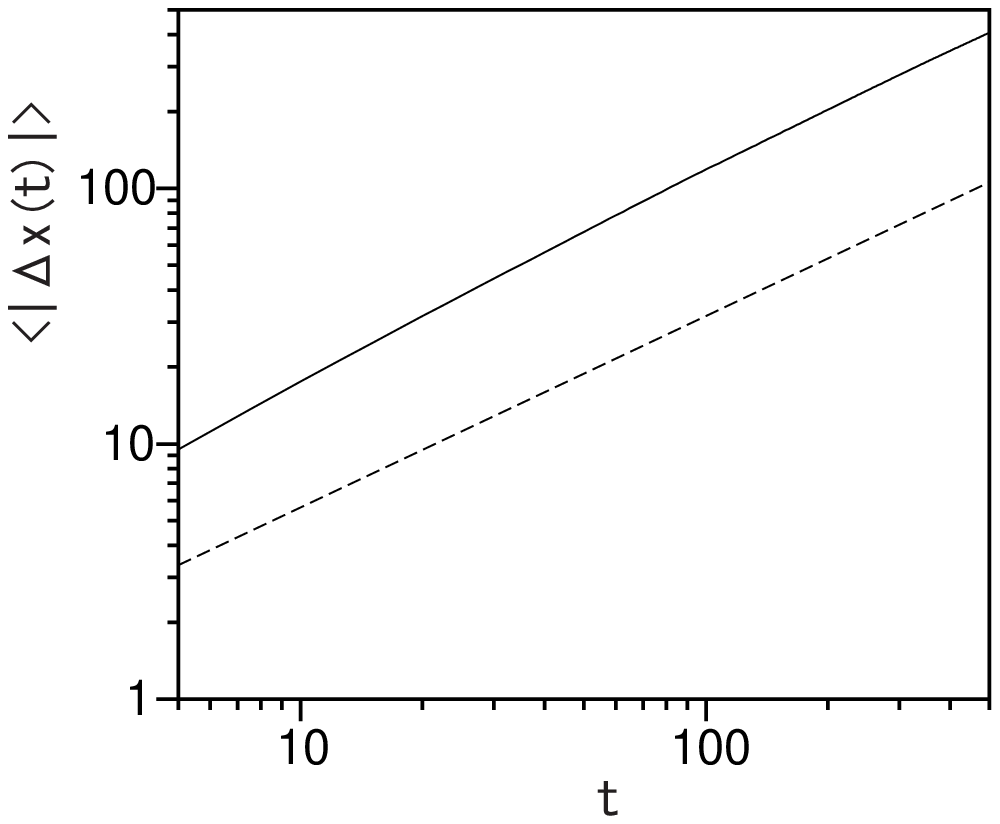}
\caption{Double logarithmic plot of $\langle |\Delta x(t)|\rangle$ vs. $t$ for Eq.~(1) and Eq.~(19) at $\gamma=1,T=1/2$ and $D=3/4$. The dashed line denotes a line with slope 3/4.} 
\label{fig:6} 
\end{center}
\end{figure}

When $D>\gamma/2$, the variance of $y$ diverges and the central limit theorem 
cannot be satisfied, and therefore the Gaussian distribution is not expected for the displacement $\Delta x$. 
For $D<\gamma$, the time correlation decays exponentially. The 
displacement $\Delta x(t)$ may be a summation of the small displacement in a time interval $\tau$ of the order of the decay constant. 
The small displacements in the time interval $\tau$ are expected to be independent stochastic variables, since the time correlation decays during the time interval. The small displacement is estimated as $\Delta x(\tau)\sim y(t)\tau$ and 
the stochastic variable $y$ obeys the stationary distribution with the power-law tail. The distribution of summation of $n$ independent stochastic variables, each of which obeys the distribution with a power-law tail, approaches the 
"stable distribution" with the same power-law tail.~\cite{rf:14} 
Our random walk is therefore considered to be a kind of Levy flight in one dimension.

For the "stable distribution" with the power-law tail of exponent $\beta$ where $\beta>3$, the summation of $n$ independent variables is expected to satisfy
\begin{equation}
\langle \{ \sum_{i=1}^{n} y_i \}^{\sigma}\rangle\sim n^{\sigma/\alpha},
\end{equation}
where $y_i$ denotes the $i$th independent stochastic variable and $\alpha=\beta-1$.~\cite{rf:14} 
In our model, the total time interval $t$ is proportional to $n$ and the following relation is expected.
\begin{equation}
\langle \Delta x(t)^{\sigma}\rangle\sim t^{\sigma/\alpha}.
\end{equation}
The average $\langle \Delta x(t)^2\rangle$ will diverge, since the variance diverges, however, the average $\langle \Delta x(t)^{\sigma}\rangle$ for $\sigma<2$ may be converged.
We have calculated the average $\langle |\Delta x(t)|\rangle$ for $\gamma=1,\;T=1/2$ and $D=3/4$. Figure  5  displays the time evolution of $x(t)$. The occasional large random jumps are characteristic of the Levy flight. 
Figure 6 displays the average  $\langle |\Delta x(t)|\rangle$ as a function of $t$ in a logarithmic scale. The average displacement obeys a power-law approximately. The exponent $\beta$ of the power-law tail for the parameter values is expected to be 7/3 and therefore $\alpha=4/3$.  The power-law evolution of $\langle |\Delta x(t)|\rangle$ is expressed as $t^{\delta}$ and the exponent $\delta$ is expected to be 3/4, which is different from the value 1/2 for the normal diffusion. The slope of the numerically obtained line is approximately 0.76, which is close to 3/4 and definitely different from 1/2.

\section{Summary}
We have shown that the Langevin model with multiplicative noise 
exhibits the probability distribution with the power-law tail, however, 
the fluctuation-dissipation relation is satisfied when the multiplicative noise is weak. When the multiplicative noise is sufficiently strong, the variance diverges and the Levy flight appears for the corresponding Brownian motion, whose  velocity obeys the Langevin equation.  
The Langevin equation with multiplicative noise may be a simple and instructive  model  in the nonequilibrium  statistical mechanics.

\end{document}